\documentclass[aps,prl, twocolumn, amsmath, amssymb, floatfix, showpacs]{revtex4-1}
\usepackage[final]{graphicx}
\usepackage{hyperref}
\usepackage{subfigure}
\usepackage{enumerate}
\graphicspath{{Figures/}}
\usepackage{color}

\begin{document}
\title{Neutron Reflectometry Studies on Magnetic Stripe Domains in Permalloy/Superconductor bilayers}
\author{Yaohua~Liu$^{1}$}\email[]{yhliu@anl.gov}  \author{M.~Iavarone$^{2}$} \author{A.~Belkin$^{1}$}  \author{G.~Karapetrov$^{1, 3}$}  \author{V.~Novosad$^{1}$}  \author{M.~Zhernenkov$^{4}$} \email[]{Current Address : Brookhaven National Laboratory, Upton, New York 11973, USA} \author{Q.~Wang$^{4}$} \author{M.~R.~Fitzsimmons$^{4}$}  \author{V.~Lauter$^{5}$}  \author{S.~G.~E.~te~Velthuis$^{1}$}  \email[]{tevelthuis@anl.gov}

\affiliation{$^{1}$Materials Science Division, Argonne National Laboratory, Argonne, Illinois 60439, USA}\affiliation{$^{2}$Department of Physics, Temple University, Philadelphia, Pennsylvania 19122, USA}\affiliation{$^{3}$Department of Physics, Drexel University, Philadelphia, Pennsylvania 19104, USA} \affiliation{$^{4}$Los Alamos National Laboratory, Los Alamos, New Mexico 87545, USA}\affiliation{$^{5}$Spallation Neutron Source Oak Ridge National Laboratory, Oak Ridge, TN 37831, USA}

\date{\today}
\begin{abstract}

We explored changes in magnetic domain structures in a magnetic layer due to the onset of the superconductivity of an adjacent superconductive layer using neutron reflectometry. Magnetic domain structures in 1~$\mu$m thick permalloy (Py) films were studied as functions of magnetic field, temperature and under the influence of the onset of superconductivity in a neighboring layer.  Bragg peaks in the off-specular scattering were observed at low fields following saturation with an in-plane field, which are attributed to the quasi-parallel magnetic stripes along the field direction. During the magnetization reversal from saturation, the stripe pattern shows increases in the period, the transverse coherence length (\textit{i.e.}, perpendicular to the stripes) and the amplitude of the out-of-plane magnetization component. The coherence length of the magnetic stripes is anisotropic in the remnant state with the longitudinal coherence length (\textit{i.e.}, along the stripes) being larger than the transverse one. The stripe period shows a weak temperature dependence between 300~K and 3~K, but no abrupt change in the period is observed when the temperature crosses the superconducting critical temperature. 
\end{abstract}

\pacs{75.25.-j, , 75.47.-m}
%75.25.-j 	Spin arrangements in magnetically ordered materials (including neutron and spin-polarized electron studies, synchrotron-source x-ray scattering, etc.)
%75.70.Cn 	Magnetic properties of interfaces (multilayers, superlattices, heterostructures)
%75.47.-m Magnetotransport phenomena; materials for magneto transport
\maketitle

\section{Introduction}

Interesting stripe patterns occur in many systems~\cite{harrison2000mechanisms}, for example, zebras, smectic liquid crystals, aligned copolymers and magnetic thin films \textit{etc}. Stripe patterns in magnetic thin films are unique because of their electronic rather than steric origin and high tunablility~\cite{murayama1966micromagnetics}. The tunability of the magnetic domain pattern is beneficial because it not only enables potential applications in logic and memory devices~\cite{allwood2005magnetic, parkin2008magnetic}, but also allows modification of the properties of layers directly coupled to the magnetic layer. For example, in artificial ferromagnetic/superconducting (FM/SC) hybrids, magnetic domain walls have been used to spatially confine the superconductivity~\cite{gillijns2005domain, aladyshkin2006thin, zhu2008altering} while magnetic domains have been used to guide and pin vortices in the adjacent superconducting layer~\cite{yang2004domain, belkin2008superconductor, vlasko2008guiding, karapetrov2009transverse, iavarone2014visualizing}.  Most studies focus on how superconductivity is influenced by the magnetic structure of the ferromagnetic layer, such as effects from stray fields~\cite{yang2004domain, belkin2008superconductor, vlasko2008guiding} or induced exchange fields~\cite{liu2012effect, liu2012exchange}, inverse proximity effect~\cite{xia2009inverse}, and induced triplet superconductivity~\cite{zhu2013unanticipated, robinson2010controlled, khaire2010observation}.  In these studies, it is typically taken for granted that the magnetic configuration of the FM layer remains intact upon the onset of superconductivity in the adjacent layer, because the energy scale associated with magnetization is typically much larger than that of superconductivity.  However, the long range effects from the interactions of FM stray fields with SC screening currents can change magnetic domain patterns. Thus, recently there is increased interest in exploring how the onset of superconductivity modifies the magnetic structures in FM/SC bilayers. Several techniques have been employed, including X-ray magnetic circular dichroism~\cite{freeland2007magnetic}, SQUID magnetometery~\cite{yashwant2012magnetic} and magneto-optical imaging techniques ~\cite{vlasko2010coupled, tamegai2011experimental, vlasko2012domain}.  Here neutron reflectometry has been used to probe changes of the magnetic stripe domains in permalloy (Py) films in Py/Nb and Py/MoGe hybrids as a function of magnetic field and temperature.  These specific hybrids are of interest as previous work has found a significant influence from the magnetic domain structure on the superconducting vortex dynamics in Py/MoGe hybrids~\cite{belkin2008superconductor, vlasko2008guiding} and on vortex formation in Py/Nb bilayers~\cite{iavarone2011imaging, bobba2014vortex}. However, whether the onset of superconductivity in turn modifies the magnetic structure has not been investigated in these systems.  Here, we have employed neutron reflectometry to determine changes in magnetic domain structures in Py films as functions of magnetic field, temperature and under the influence of the onset of superconductivity in a neighboring Nb (or MoGe) layer. 
\section{Experimental Techniques}

\begin{figure}[t]
	\centering
		\includegraphics[width=0.36\textwidth]{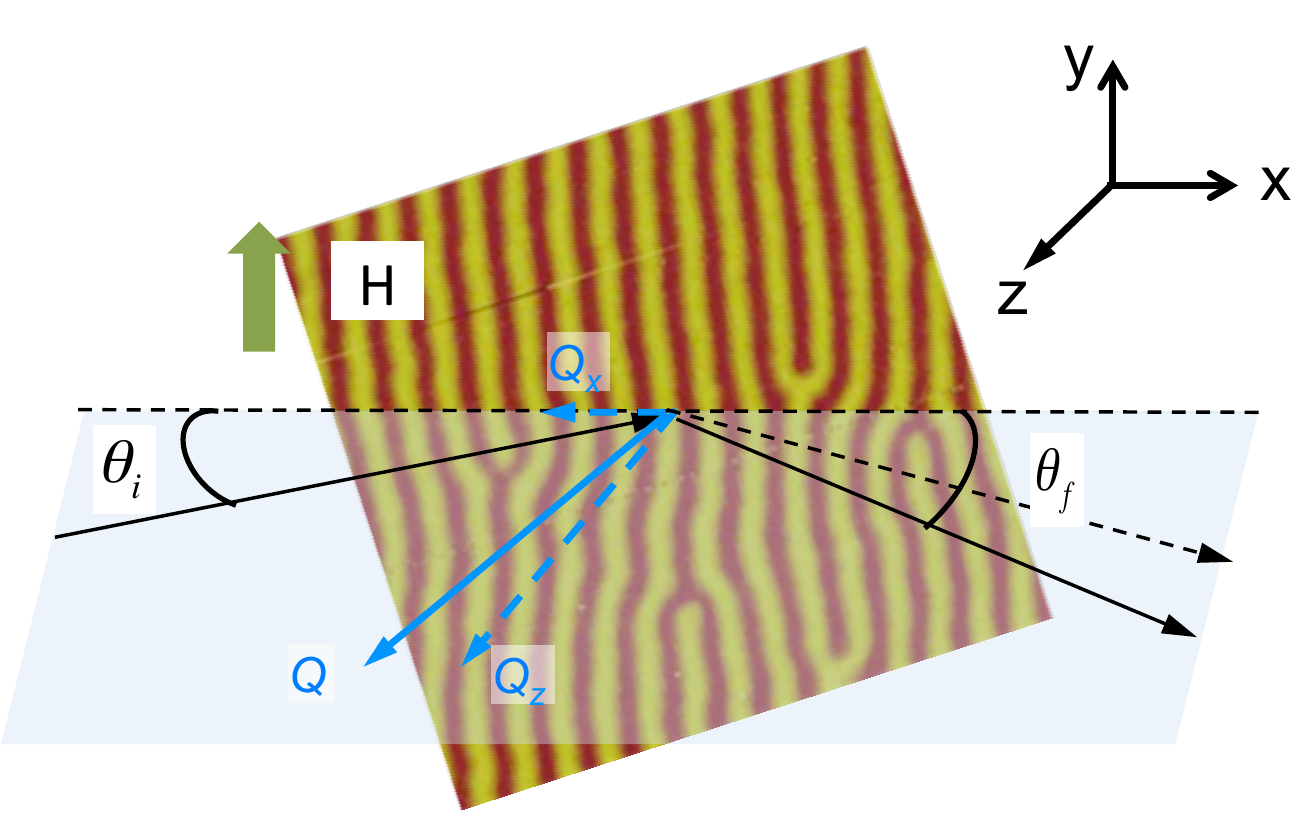}
	\caption{\label{Fig:Offspec}(Color online) The scattering geometry during the  neutron reflectometry experiments. The path of the neutron beam is indicated by thick black arrows. The wave vector transfer $Q$ has a specular component $Q_z = 4 \pi sin\theta_i /\lambda$ and an off-specular component $Q_x \approx \frac{2 \pi}{\lambda} (cos\theta_f - cos\theta_i)$, within the reflection plane (light blue parallelogram lying in the $xz$-plane). Here $\lambda$ is the neutron wavelength. The square image shows a magnetic stripe domain structure of a thick Py film probed by magnetic force microscopy at room temperature. An in-plane magnetic field (green arrow) is used to align the domains along the $y$-axis, \textit{i.e.}, perpendicular to the reflection plane. The stripe domains in the Py film provide a 1D quasi-periodic scattering potential for neutrons and effectively scatter the neutrons constructively in an off-specular direction.}
\end{figure}

Neutron reflectometry allows to probe variations of the film properties as a function of depth~\cite{FelcherRSI1987} and to determine in-plane correlations with sensitivity to the orientation of the magnetization vector~\cite{Fermon1999offspec, felcher2001perspectives, lauter2000magnetic}. For the specular neutron reflectivity, the incident angle of the neutron beam relative to the sample surface is equal to the reflected angle ($\theta_i = \theta_f$) and thus the wavevector transfer $Q$ is equal to $Q_z$ (see Fig.~\ref{Fig:Offspec}). Hence, specular reflectivity is determined by the depth profiles of both the chemical structure and the magnetization along the film stacking direction $z$~\cite{FelcherRSI1987,  MajkrzakPhysB1996}. Off-specularly reflected intensity ($\theta_i \neq \theta_f$), originating from in-plane correlations along $x$,  has a non-zero in-plane component of the wavevector transfer ($Q_x  \neq 0$)~\cite{zabel2007polarized}.  Correlations along $y$ result in scattering out of the reflection plane ($Q_y \neq 0$), which is typically referred to as Grazing Incidence scattering (GIS, either diffraction or small angle scattering)~\cite{ott2007neutron, zabel2007polarized}. As the magnetic scattering cross-section is zero when $Q || M$,  the specular polarized neutron reflectivity can only determine the magnitude and orientation of the magnetization components within the plane ($M_x, M_y$), however off-specular and grazing incidence scattering are sensitive to modulations of $M_z$. Further sensitivity to the direction of the magnetization is obtained by comparing spin-flip and non-spin flip intensities. Note that spin-flip intensities are solely determined by magnetization components that are perpendicular to the neutron spin quantization axis. This axis, \textit{i.e.} polarization direction, is typically determined by the direction of the applied magnetic field. 

At remanence after in-plane saturation, the domain structure of a thick ($\sim$1~$\mu$m) Py film consists of the quasi-periodic stripes with spatially oscillating in- and out-of-plane components of magnetization, as illustrated by a magnetic force microscopy (MFM) image in Fig.~\ref{Fig:Offspec}~\cite{murayama1966micromagnetics, spain1963dense, saito1964new}. This pattern, forming above a certain critical film thickness, is the result of a growth-induced perpendicular anisotropy~\cite{youssef2004thickness}. The stripe domains form predominately along the original saturating field direction, yet contain both disclination and dislocation defects. These defects are important for stripe ordering and limit the coherence length of the domain pattern~\cite{harrison2000mechanisms}.  The contrast in the MFM image of a Py film in Fig.~\ref{Fig:Offspec} arises from the variation of the out-of-plane magnetization component that gives rise to a modulation of the scattering potential transverse to the stripe direction, thus can effectively scatter neutrons~\cite{halpern1939magnetic}.  In an analogy, the stripe pattern behaves like a one dimensional micrograting for neutrons, and gives rise to Bragg scattering peaks in off-specular or grazing incidence directions, the positions of which depend on the period of the grating. Previously, GIS has been used to study the out-of-plane magnetization of magnetic stripe domains~\cite{Fermon1999offspec, Ott2006FeFeNGISANS, ott2007neutron} and magnetic vortex cores~\cite{Schuller2009VortexCore} with correlations lengths below 100~nm. Here, we evaluated  the off-specular scattering using neutron reflectometers, which characterizes larger correlations ($>$ 500 nm~\cite{ott2007neutron}) and has so far only been used to probe domains with an in-plane magnetization or structural correlations of the surface~\cite{lauter2000magnetic, lauter2002transverse}.  In a typical neutron reflectometer, the beam is relatively divergent along $y$ to maximize the intensity, rather than being highly collimated in all directions as required for resolving grazing incidence scattering. 

\section{Samples and Instruments}

The nominal structures of the two studied samples are Nb (100~nm)/SiO$_x$ (10~nm )/Py (1~$\mu$m )//native SiO$_x$/Si substrate, and Mo$_{79}$Ge$_{21}$ (40~nm)/SiO$_{x}$ (20~nm )/Py (1~$\mu$m)//native SiO$_{x}$/Si substrate, which are referred as the Nb/Py and MoGe/Py samples below, respectively. The Py (Ni$_{79}$Fe$_{21}$) and MoGe layers were grown by dc sputtering at room temperature at a base pressure of 1.5 $\times$ 10$^{-7}$ Torr. The Nb films were grown in a dedicated dc sputtering system at a pressure of 5.8 $\times$ 10$^{-9}$ Torr. A SiO$_{x}$ layer was deposited between the FM and SC layers in order to suppress the proximity effect. The superconducting critical temperatures ($T_{C}$) are 6.2~K~\cite{belkin2008superconductor} and 9.0~K~\cite{iavarone2011imaging} for the MoGe/Py and Nb/Py films, respectively.  Superconducting MoGe and Nb films have quite different penetration depths, thus they might display different screening effects on the stray fields from the underlying Py films.

%penetration length:  
%Nb at 7.2 K, 61/(1-(7.2/9)**4)**0.5 = 80 nm

Neutron reflectivity experiments were performed using the Asterix reflectometer at the Lujan Neutron Scattering Center at Los Alamos National Laboratory and the Magnetism Reflectometer beamline at the Spallation Neutron Source at Oak Ridge National Laboratory. Both instruments use the time-of-flight technique and have position sensitive detectors, allowing the reflected intensity to be determined for a range of $Q_z$ and $Q_x$ values, respectively, with one setting of the incident angle $\theta_i$. The wavelength bands used were 4~\AA~to 13~\AA~on Asterix, and 3~\AA~to 6~\AA~on the Magnetism Reflectometer. Both instruments provide an incident polarized beam, however the experiments aimed at measuring the off-specular intensities were performed with unpolarized neutrons, unless noted otherwise.

\section{Results}

\subsection{Field Dependence}

\begin{figure}[t]
	\centering
		\includegraphics[width=0.5\textwidth]{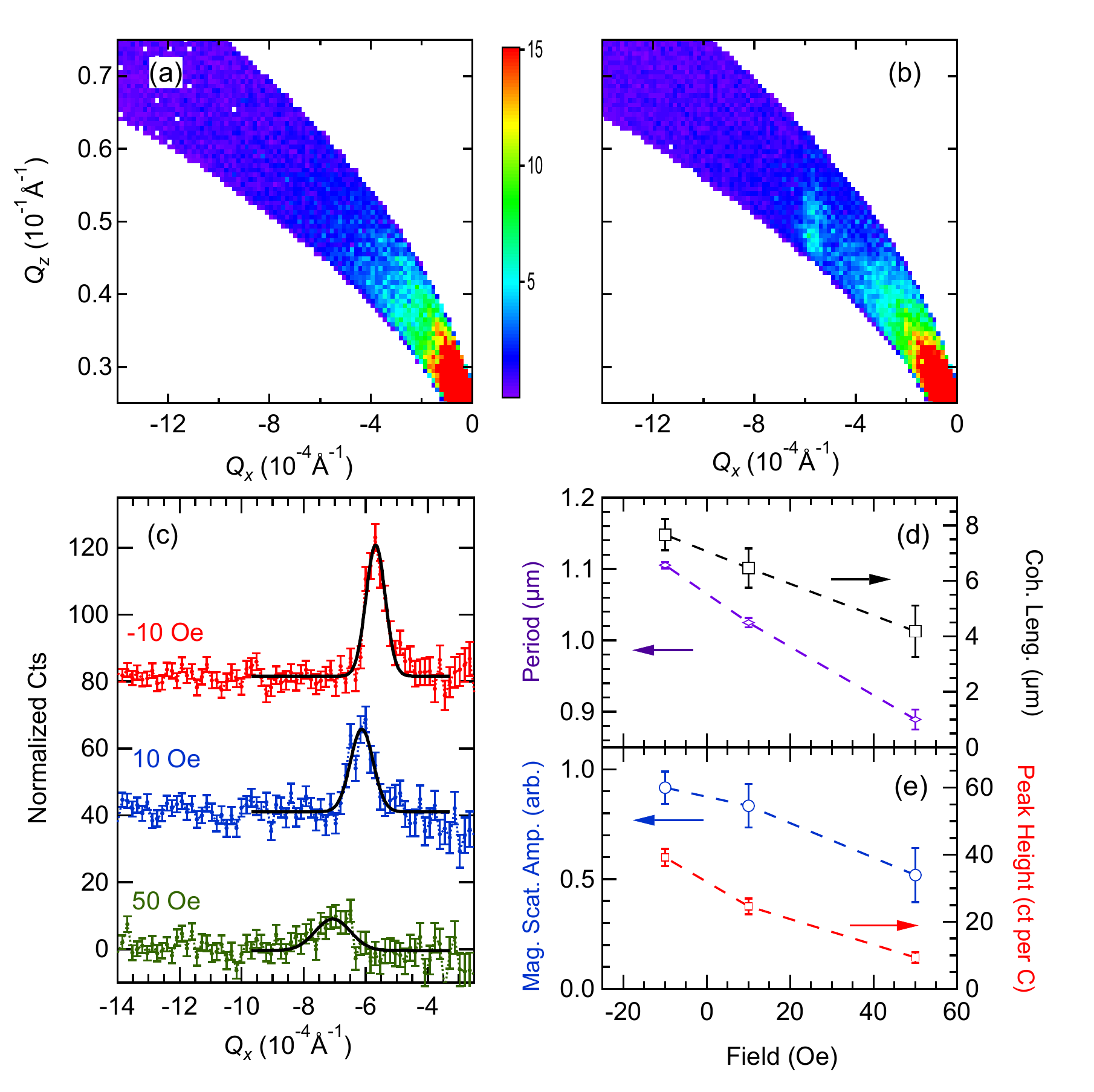}
	\caption{\label{Fig:Nb_Hscan}(Color online)  The reciprocal space intensity maps taken with in-plane field $H_y = $ 5~kOe (a) and $H_y = $ -10~Oe (b) after saturation in 5~kOe from a Nb/Py sample. An off-specular Bragg peak shows up around $Q_x$ of $\sim 6 \times 10^{-4}$~\AA$^{-1}$ at low fields.  (c) The off-specular scattering intensities integrated along the $Q_z$ direction. The 5~kOe data have been subtracted from the low-field data. The peak position and the FWHM of the off-specular peaks are obtained by fitting the data with a Gaussian function. (d) and (e) The period, the transverse coherence length, the peak height and relative magnetic scattering amplitude as functions of magnetic field. } 
\end{figure}

Experiments were initially performed above $T_{C}$ of the superconductors and at various applied magnetic fields in order to establish the sensitivity of the neutron reflectometry to the magnetic stripe domains within the Py layer. Above $T_{C}$ the characteristics of Py determined for either sample can be considered to be typical for both, because their nominal thicknesses are the same and only the top superconducting layers are different. Data was first collected at room temperature with an in-plane magnetic field $H_y = 5$~kOe, which is far above the saturation field, then consecutively at $H_y =$ 50~Oe, 10~Oe, and -10~Oe. This field history is expected to align the magnetic stripes to the polarization direction of the neutron beam, as shown in Fig.~\ref{Fig:Offspec}. Figures~\ref{Fig:Nb_Hscan}(a) and~\ref{Fig:Nb_Hscan}(b) show the reciprocal space intensity maps from the Nb/Py sample at 5~kOe and -10~Oe, respectively, measured on the Magnetism Reflectometer with $\theta_i=0.4^\circ$. At low fields, there is indeed off-specular scattering near $Q_x$ of $\sim 6 \times 10^{-3}$~\AA$^{-1}$ that is absent at 5~kOe, which indicates that the off-specular scattering is related to the magnetic domains within the plane of the film. To enable determination of the peak position and width of the off-specular scattering more clearly, the measured scattering intensities were integrated along $Q_{z}$ and the 5~kOe data was subtracted as a background from the low field data, effectively removing any contribution from the chemical structure.  The results, plotted as a function of $Q_x$, are shown in Fig.~\ref{Fig:Nb_Hscan}(c). From these curves, the peak position and width of the off-specular peaks were determined via fitting with a Gaussian function, from which subsequently the period and the transverse (\textit{i.e.}, perpendicular to the stripes) coherence length are calculated, as shown in Fig.~\ref{Fig:Nb_Hscan}(d). The period can be determined with a relative uncertainty of about 0.5-1.5\%, depending on the field. 

The measured period is around 1 $\mu$m, which is close to the value being calculated based on the film thickness, magnetic anisotropy, exchange constant, and saturation magnetization~\cite{murayama1966micromagnetics}. In the range of 50~Oe to -10~Oe during the descending field scan, the period increases about 3.3~nm per Oe (a relative change rate of $\sim$ 0.3\% per Oe).  A similar trend of the field dependence has been observed by magnetic force microscopy and reproduced by two-dimensional micromagnetic simulations~\cite{talbi2010magnetic}. Since neutron scattering is a non-local probe, it allows not only to determine the average period of the stripes, but also to extract a statistical average of the coherence lengths of the stripe domains. The transverse coherence length of the stripe pattern is estimated by the Scherrer equation, $L \sim 0.89\times 2 \pi / \sqrt{\beta_B^2 - \beta_0^2}$, where $\beta_B$ and $ \beta_0$ are the full width at half maximum of the off-specular Bragg peak and the instrument resolution, respectively. The instrument resolution of $3 \times 10^{-6}$~\AA$^{-1}$ was estimated by the peak width of the specular reflection along  $Q_x$ at 5~kOe.  The obtained values of $L$ range from $4~\mu$m to $8~\mu$m. Note that this coherence length is not limited by that of the neutron source, which is about $25-50~\mu$m along the transversal direction for this experiment~\cite{zabel2007polarized}. 

Figure~\ref{Fig:Nb_Hscan}(e) shows the peak height and the calculated magnetic scattering amplitude, which increase as the field decreases from saturation. In the first order Born approximation, the Bragg peak intensity $I \propto N^2*a^2$, where $N$ is the number of the scattering objects within the coherence length and  $a$ is the scattering amplitude of each object.  Here the scattering amplitude is determined by the amplitude of the out-of-plane magnetization modulation. Thus, the results suggest an increase in the amplitude of the out-of-plane magnetization at low fields, which is consistent with a decrease in the in-plane magnetization at low fields, as observed on similar samples~\cite{vlasko2008guiding}.

\begin{figure}[t]
	\centering
		\includegraphics[width=0.36\textwidth]{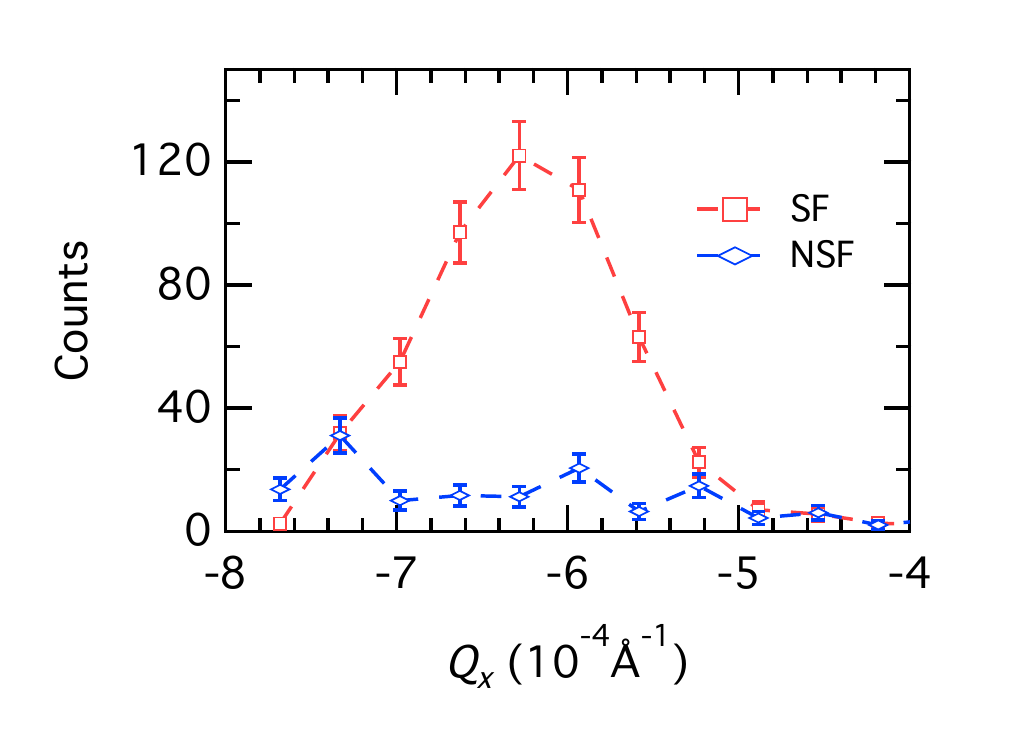}
		\caption{\label{Fig:MoGe_SF}(Color online) Polarization analysis on the off-specular Bragg peaks from the MoGe/Py sample. An off-specular Bragg peak is only seen in the spin-flip cross section, indicating its magnetic origin. } 
\end{figure}

To verify that out-of-plane modulations of the magnetization are at the origin of the off-specular Bragg peak, polarized neutron reflectometry with polarization analysis was performed. The initial neutron polarization was along the $y$ direction, as determined by the applied field direction during the measurements.Off-specular scattering ($Q_x \neq 0$) resulting from variations in $M_y$ would be non-spin flip, while variations in $M_z$ would be spin-flip. These experiments were performed with the MoGe/Py sample on Asterix with an incident angle of 0.8$^\circ$. Figure~\ref{Fig:MoGe_SF} shows both the spin-flip and non-spin flip data collected at 7.5~K in a 8.5~Oe in-plane field after saturation.  The off-specular Bragg peak is only visible in the spin-flip channel, which confirms modulations in the out-of-plane magnetization ($M_z$) are solely at the origin of this intensity, and $M_y$ does not vary significantly between domains.

\subsection{Domain Pattern Anisotropy}

\begin{figure}[t]
	\centering
		\includegraphics[width=0.5\textwidth]{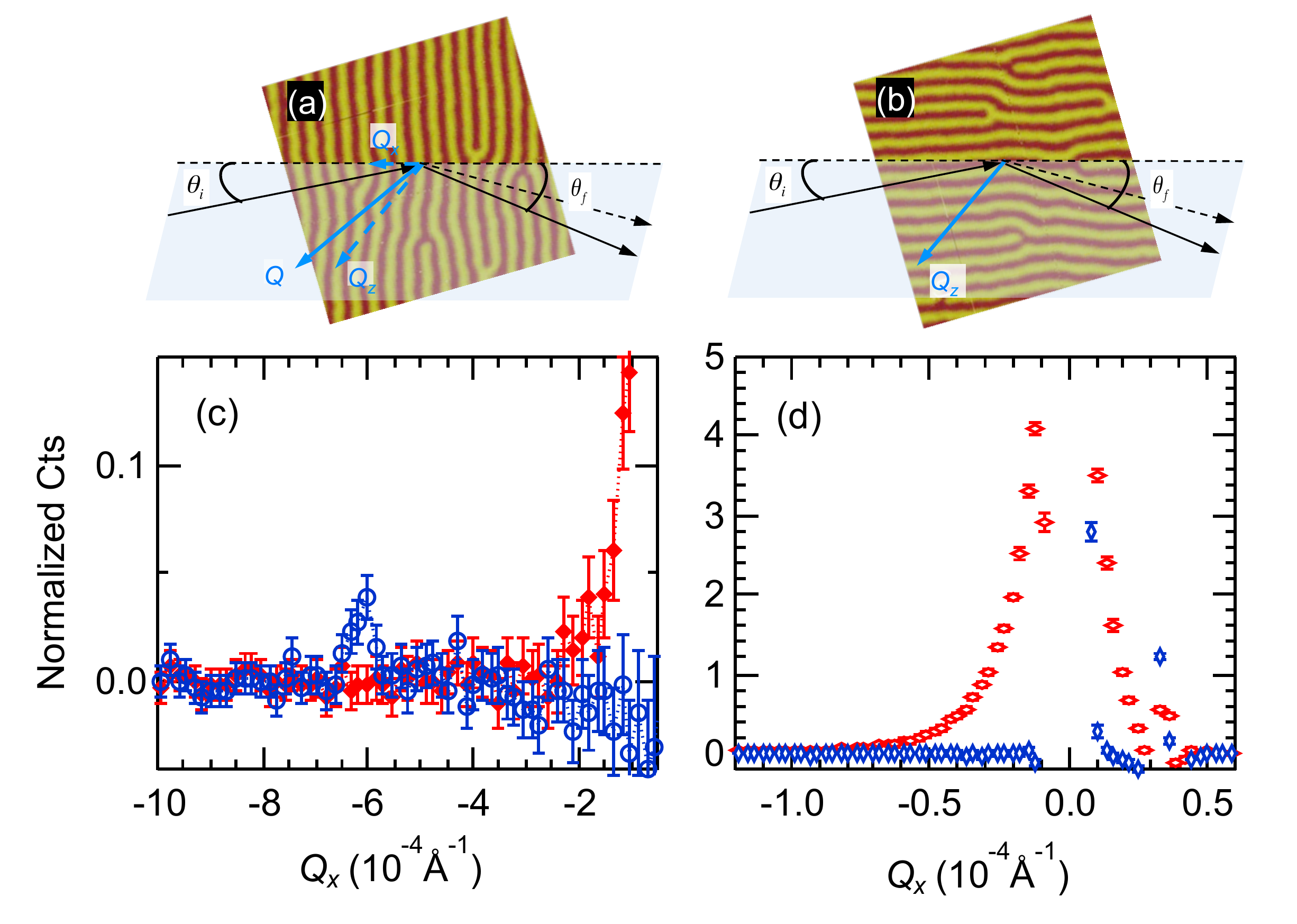}
	\caption{\label{Fig:Py_HRot}(Color online) (a) and (b) show schematically two scattering geometries,  where the data were collected at zero field right after saturation and after rotating the sample by 90$^\circ$ along the film surface normal, respectively. (c) and (d) show the off-specular scattering intensity from the Nb/Py sample in both geometries around the off-specular Bragg peak position and the specular position, respectively. The blue open circles and the red solid diamonds show the data collected corresponding to the case (a) and (b), respectively. The off-specular Bragg peak around $Q_x ~\sim 6 \times 10 ^{-4}$~\AA$^{-1}$ is only visible in the first case.  At the same time, there is obvious diffuse scattering near $Q_x = 0$  in the latter case.  A  longitudinal coherence length of 13~$\mu$m is estimated from the width of the diffuse scattering. } 
\end{figure}

The presence of stripes, rather than maze-like magnetic domains, and the longitudinal (i.e. along the stripes) coherence length were determined by comparing measurements of the Py/Nb sample in two different orientations with respect to the neutron beam. The experiments were performed at room temperature on the Magnetism Reflectometer with an incident angle of 0.4$^\circ$. In the first configuration, the data were collected after a saturating field applied along $y$ was reduced to zero; subsequently for the second configuration, the sample was rotated around the surface normal by 90$^\circ$. Figure~\ref{Fig:Py_HRot} shows the off-specular intensity in the two configurations after subtracting the 5~kOe data, as was done for Fig.~\ref{Fig:Nb_Hscan}(c). The off-specular Bragg peak around $Q_x ~\sim 6 \times 10 ^{-4}$~\AA$^{-1}$ is visible in the first case but is absent after the sample is rotated. This means the magnetic domain pattern is anisotropic, as expected for stripes, and rules out maze-like magnetic domains. In the 90$^\circ$ sample orientation, one would expect to see grazing incidence scattering with Bragg peaks at $Q_y~\sim 6 \times 10 ^{-4}$~\AA$^{-1}$. However, with the experimental setup used, that would correspond to a scattering angle along $y$ that is, at best, five thousand times smaller than the one detected along the $x$ direction. Due to the relatively large divergence of the neutron beam in the $y$ direction and limited angular resolution of the detector, such a small change in angle cannot be resolved, and the scattering is indistinguishable from the specular reflectivity. 

More interestingly, there is a second difference between the scattered intensities for the two orientations of the sample close to $Q_x = 0$. There is a significant amount of diffuse scattering around the specular reflection after the sample is rotated by 90$^\circ$, \text{i.e.} when the magnetic stripes are parallel to the reflection plane, as shown in Fig.~\ref{Fig:Py_HRot}b.  From the width of the diffuse scattering, the longitudinal coherence length of the magnetic structure along the stripe direction is estimated to be 13~$\mu$m,  which is roughly twice the transverse one. The origin of the anisotropy of the coherence length is very intriguing because for a perfect and infinitely large stripe pattern, the coherence length is isotropic and infinite in any direction. The origin of the anisotropy is likely caused by the magnetic structural defects. To our knowledge, the coherence length anisotropy of magnetic domain patterns has not been discussed in the literature, but it is certainly worth future investigation.

\begin{figure}[t]
	\centering
		\includegraphics[width=0.4\textwidth]{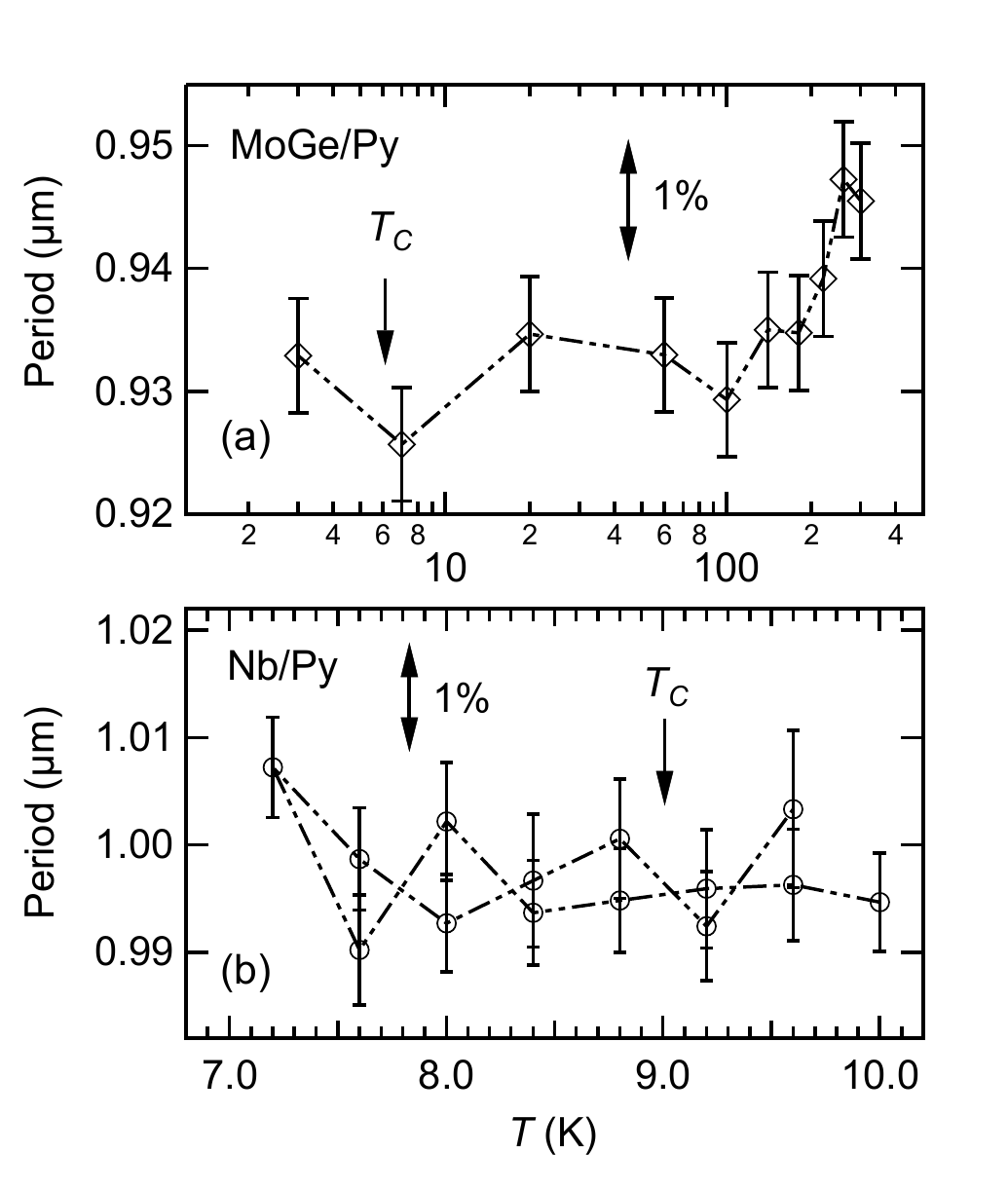}
	\caption{\label{Fig:OFF_TScan}(a) Temperature dependence of the off-specular Bragg peak position from the MoGe/Py sample during cooling.  The data were collected at 8.5~Oe after saturation in an in-plane magnetic field of 540~Oe at 300~K. There is a small change of the period of $\sim$1.5\% between 300~K and 100~K, but no noticeable change below 100~K.  (b) Temperature dependence of the off-specular Bragg peak position from the Nb/Py sample.  The $T_{C}$ of the MoGe/Py and Nb/Py films are indicated by arrows. In both cases, there is no abrupt change of the off-specular Bragg peak position when the temperature crossed the $T_{C}$.} 	
\end{figure}

\subsection{Temperature Dependence}
The temperature dependence of the stripe domain period was investigated in both the MoGe/Py and Nb/Py samples. Figure~\ref{Fig:OFF_TScan}(a) shows the temperature dependence of the off-specular Bragg peak position from the MoGe/Py sample. These experiments were performed using Asterix, with an incident angle of 0.8$^\circ$.  The sample was first saturated in an in-plane magnetic field of 540~Oe at 300~K and then the field was reduced to 8.5~Oe. Data were collected at each temperature during cooling from 300~K to 3.2~K. The period shows a weak temperature dependence.  It gradually decreases about 1.5\% when the temperature decreases from 300~K to 100~K, then barely changes below 100~K.  There is no indication of an abrupt change in the period when the temperature crosses the $T_{C}$ of MoGe, which is 6.2~K. The Nb/Py sample was studied at the Magnetism Reflectometer in the vicinity of $T_{C}$, but now with fine temperature steps, as shown in Fig.~\ref{Fig:OFF_TScan}(b). The experiments started from 10~K, and then cooled down to 7.2~K, the lowest temperature achievable during the experiments, and then warmed back to 9.6~K, with a step size of  0.4~K. Since the temperature range was small during the experiments, all instrumental parameters, except the temperature, were kept fixed to avoid potential uncertainty from re-alignment between consecutive runs.  Similar to the results from the MoGe sample, there is no significant or abrupt change of the off-specular Bragg peak position when the temperature crosses the $T_{C}$ of Nb, which is 9.0~K.  From these two experiments it can be concluded that there is a slight decrease in the period of the stripe pattern in Py when temperature changes from 300~K to 100~K, and the upper limit of the relative change is $\sim$1\% when the temperature cross the $T_{C}$ in the two samples. 

\begin{figure}[t]
	\centering
		\includegraphics[width=0.5\textwidth]{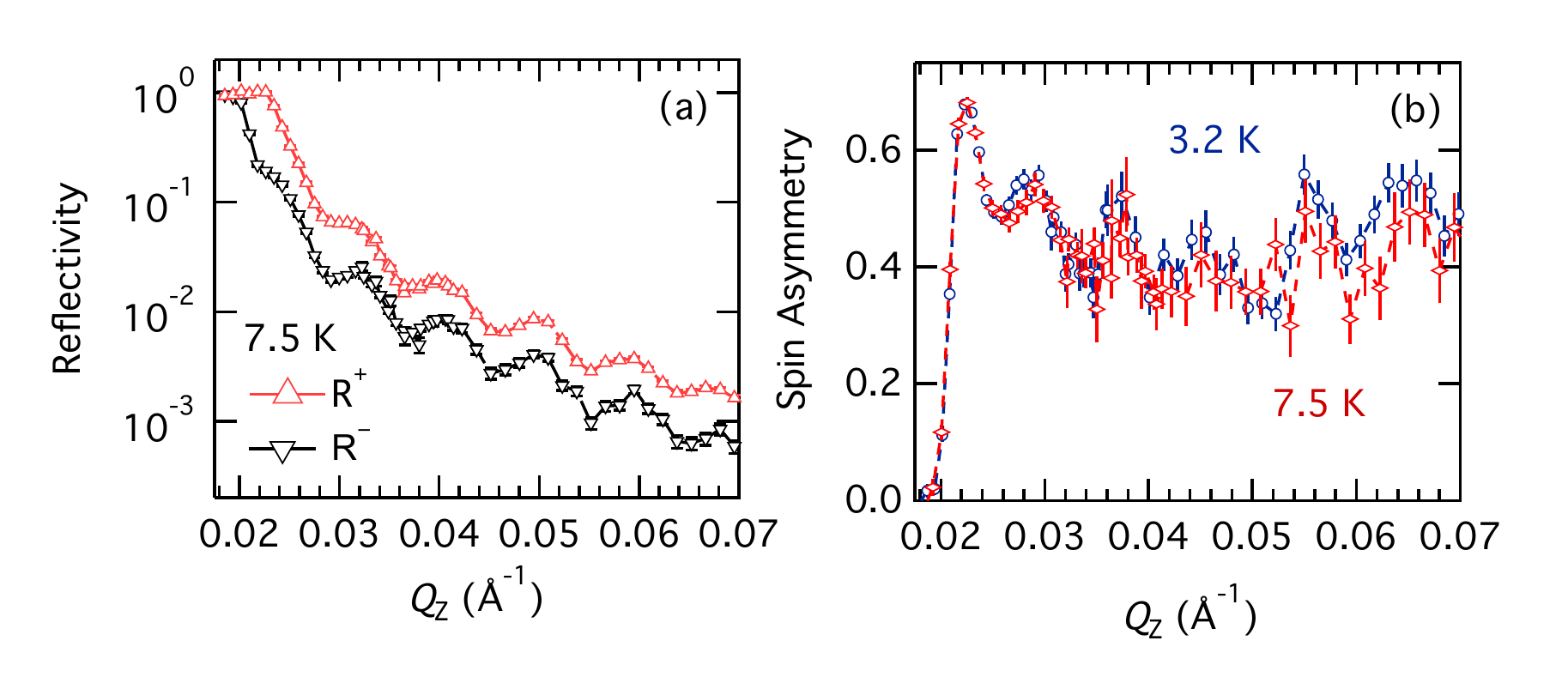}
	\caption{\label{Fig:Spec_ASY} (a) Specular polarized neutron reflectivity measured at 7.5~K; (b) Spin asymmetry above (7.5~K) and below (3.2~K) the superconducting transition temperature of the MoGe/Py sample.} 	
\end{figure}

It is worth noting that in the analysis of the off-specular scattering, the intensity along $Q_{z}$ was integrated, thereby eliminating any depth sensitivity. Therefore, these results do not rule out the possibility that the domain pattern at the top surface Py layer, closest to the superconductor, changes when the temperature crosses $T_{C}$. There is in fact a slight difference between the specular polarized neutron reflectivities above and below $T_C$ of the MoGe/Py sample. Figure~\ref{Fig:Spec_ASY}(a) shows the specular reflectivities $R^{+}$ and  $R^{-}$, measured with polarization of the incident neutrons parallel or antiparallel to the applied field, respectively at 7.5~K. Figure \ref{Fig:Spec_ASY}(b) shows the spin asymmetry, which is defined as,  $\frac{R^{+} - R{-}}{R^{+} + R^{-}}$, at 7.5~K and 3.2~K. The spin asymmetry is slightly different for the two temperatures, indicating the in-plane magnetic induction of the sample has slightly changed upon crossing $T_{C}$. The specular reflectivity is affected by the depth profile of the in-plane components of $B$, thus the change in the spin asymmetry could be due to a change in the magnetic domain pattern, and/or a change in the profile of $B$ due to field penetration and vortices in MoGe. For this experiment, quantitative analysis of the specular data is very challenging as it would require a full dynamic scattering theory to take into account contributions to the magnetic induction from the domains, stray fields at the surface of the domains ~\cite{Fermon1999offspec},  as well as variations of $B$ in MoGe below $T_{C}$. Due to these complications and the small change in the spin asymmetry, such an analysis would unlikely be able to determine the magnetization change, thus was not attempted. Overall, the off-specular results show that the long range effects, associated with interactions of the FM stray fields with the SC screening currents, barely changes the average magnetization domain patterns in the Py films.

\section{Summary}
In summary, off-specular scattering resulting from the out-of-plane magnetization components of magnetic stripe domains in thick permalloy films in ferromagnetic/superconducting hybrid structures has been successfully observed in neutron reflectometry experiments, which illustrates the feasibility of such studies. The stripe pattern in Py is found to be anisotropic in the remnant state, with the longitudinal coherence length (\textit{i.e.} along the stripes) being larger than the transverse one (\textit{i.e.} perpendicular to the stripes). The period, the transverse coherence length, and modulation amplitude of the out-of-plane magnetization component depend strongly on the field amplitude, consistent with expectations. A weak temperature dependence of the period is observed between 300~K and 100~K, however a significant change of the stripe pattern could not be observed when the temperature crosses the superconducting critical temperature $T_{C}$ of a neighboring superconducting layer. Therefore, the long range effects, associated with interactions of the FM stray fields with the SC screening currents, barely modify the average magnetization configurations in these samples, although changes at the surface of Py cannot be ruled out.  

\begin{acknowledgments}
We thank Dr. R. Osborn (ANL) for insightful discussions, and Dr. H. Ambaye and R. Goyette (ORNL) for assistance during the neutron experiments. Work at Argonne National Laboratory (YL, AB, GK, and StV) was supported by the U.S. Department of Energy, Office of Science, Basic Energy Sciences, Materials Sciences and Engineering Division. Work at Temple University was supported by the U.S. Department of Energy, Office of Basic Energy Sciences, Division of Materials Sciences and Engineering under Award DE-SC0004556. Neutron scattering experiments at the Lujan Center for Neutron Scattering, Los Alamos National Laboratory were supported by DOE, Office of Science, BES. Los Alamos National Laboratory is operated by Los Alamos National Security LLC under DOE Contract DE-AC52-06NA25396. Neutron experiments conducted at ORNL's Spallation Neutron Source was sponsored by the Scientific User Facilities Division, Office of Basic Energy Sciences, US Department of Energy. 

\end{acknowledgments}
\bibliography{Py}	
\end{document}